\documentclass[a4paper,twocolumn,english,aps,prl,reprint,superscriptaddress,showpacs,showkeys,longbibliography]{revtex4-1}
\usepackage[latin9]{inputenc}
\usepackage{babel}
\usepackage{amsmath}
\usepackage{amssymb}
\usepackage{graphicx}
\usepackage[colorlinks=true, linkcolor=red, citecolor=blue, urlcolor=blue]{hyperref}
\usepackage[usenames,dvipsnames]{color}
\usepackage{braket}
\usepackage{float}
\usepackage{comment}

\definecolor{mygreen}{rgb}{0,0.5,0}
\definecolor{myblue}{rgb}{0,0,0.75}
\definecolor{mymagenta}{cmyk}{0,1,0,0.12}

\def\wt#1{\inner(#1)}
\def\inner(#1,#2,#3,#4,#5,#6){\ensuremath\left(\begin{array}{ccc} #1 & #2 & #3 \\ #4 & #5 & #6 \end{array}\right)}

\def\ws#1{\innerv(#1)}
\def\innerv(#1,#2,#3,#4,#5,#6){\ensuremath\left\{\begin{array}{ccc} #1 & #2 & #3 \\ #4 & #5 & #6 \end{array}\right\}}

\newcommand{\minus}{
  \setbox0=\hbox{-}
  \vcenter{
    \hrule width\wd0 height \the\fontdimen8\textfont3
  }%
}

\def\pl{\texttt{+}}
\def\mi{\texttt{-}}

\newcommand{\usa}{\pl_a}
\newcommand{\us}{\pl_s}

\newcommand{\dsa}{\mi_a}
\newcommand{\ds}{\mi_s}
\newcommand{\Uu}{U_{\pl\pl}^{(s)}}
\newcommand{\Ua}{U_{\pl\pl}^{(a)}}

\newcommand{\ssq}{\scalebox{0.5}{$\square$}}
\newcommand{\ls}{\tilde \sigma}
\newcommand{\Rvec}{{\bf R}}
\newcommand{\laserlabel}{\lambda}

\begin{document}

\title{A Coherent Quantum Annealer with Rydberg Atoms}

\author{A. W. Glaetzle}
\thanks{These two authors contributed equally}
\affiliation{Institute for Quantum Optics and Quantum Information of the Austrian
Academy of Sciences, A-6020 Innsbruck, Austria}
\affiliation{Institute for Theoretical Physics, University of Innsbruck, A-6020
Innsbruck, Austria}

\author{R. M. W. van Bijnen}
\thanks{These two authors contributed equally}
\affiliation{Institute for Quantum Optics and Quantum Information of the Austrian
Academy of Sciences, A-6020 Innsbruck, Austria}
\affiliation{Institute for Theoretical Physics, University of Innsbruck, A-6020
Innsbruck, Austria}

\author{P. Zoller}
\affiliation{Institute for Quantum Optics and Quantum Information of the Austrian
Academy of Sciences, A-6020 Innsbruck, Austria}
\affiliation{Institute for Theoretical Physics, University of Innsbruck, A-6020
Innsbruck, Austria}

\author{W. Lechner}
\email{wolfgang.lechner@uibk.ac.at}
\affiliation{Institute for Quantum Optics and Quantum Information of the Austrian
Academy of Sciences, A-6020 Innsbruck, Austria}
\affiliation{Institute for Theoretical Physics, University of Innsbruck, A-6020
Innsbruck, Austria}

\date{\today}

\begin{abstract}
There is a significant ongoing effort in realizing quantum annealing with different physical platforms. The challenge is to achieve a fully programmable quantum device featuring coherent adiabatic quantum dynamics. Here we show that combining the well-developed quantum simulation toolbox for Rydberg atoms with the recently proposed Lechner-Hauke-Zoller~(LHZ) architecture allows one to build a prototype for a coherent adiabatic quantum computer with all-to-all Ising interactions and, therefore, a novel platform for quantum annealing. In LHZ a infinite-range spin-glass is mapped onto the low energy subspace of a spin-1/2 lattice gauge model with quasi-local 4-body parity constraints. This spin model can be emulated in a natural way with Rubidium and Cesium atoms in a bipartite optical lattice involving laser-dressed Rydberg-Rydberg interactions, which  are several orders of magnitude larger than the relevant decoherence rates. This makes the exploration of coherent quantum enhanced optimization protocols accessible with state-of-the-art atomic physics experiments.
\end{abstract}



\maketitle

Quantum annealing is a quantum computing paradigm with the aim to solve generic optimization problems \cite{Nishimori1998,Farhi2000,Boixo2014,Boixo2016}, where the cost function corresponds to the energy of an infinite-range Ising spin glass~\cite{LUCAS2014}. Finding the optimal solution of the problem is thus equivalent to determining the ground state of the spin glass. In quantum annealing, this task is accomplished by adiabatic passage of a system of $N$ spins in the instantaneous ground state of a Hamiltonian (denoted {\it logical spin model}) of the form 
\begin{equation}
\tilde H_t\!=\! \tilde A_t \!\sum_{\nu=1}^N \tilde a_\nu\ls_x^{(\nu)} \!\!+\!\tilde B_t\!\!\left[\sum_{\nu=1}^{N} \tilde h_\nu  \ls_z^{(\nu)} \!+\!\sum_{\nu<\mu}^{N}\!\tilde J_{\mu\nu} \ls_z^{(\nu)} \ls_z^{(\mu)}\!\right]\!.
\label{eq:Hlogic}
\end{equation}
Here $\ls_{\{x,y,z\} }$ are Pauli spin operators, and scheduling functions $\tilde A_t$ and $\tilde B_t$ deform $\tilde H_t$ from a trivial initial Hamiltonian with $(\tilde A_0, \tilde B_0) = (1,0)$ and transverse local fields $\tilde a_\nu$, into the spin glass Hamiltonian with $(\tilde A_1, \tilde B_1) = (0,1)$, where the optimization problem is encoded in  local fields $\tilde h_\nu$ and infinite-range interactions $\tilde J_{\nu\mu}$~\cite{LUCAS2014}. Note that implementing Eq.~\eqref{eq:Hlogic} requires individually programmable long-range interactions $\tilde J_{\mu\nu}$, which is in contradiction to polynomially decaying interactions in cold atoms and molecule setups.

\begin{figure}[tb]
\centering
\includegraphics[width= \columnwidth]{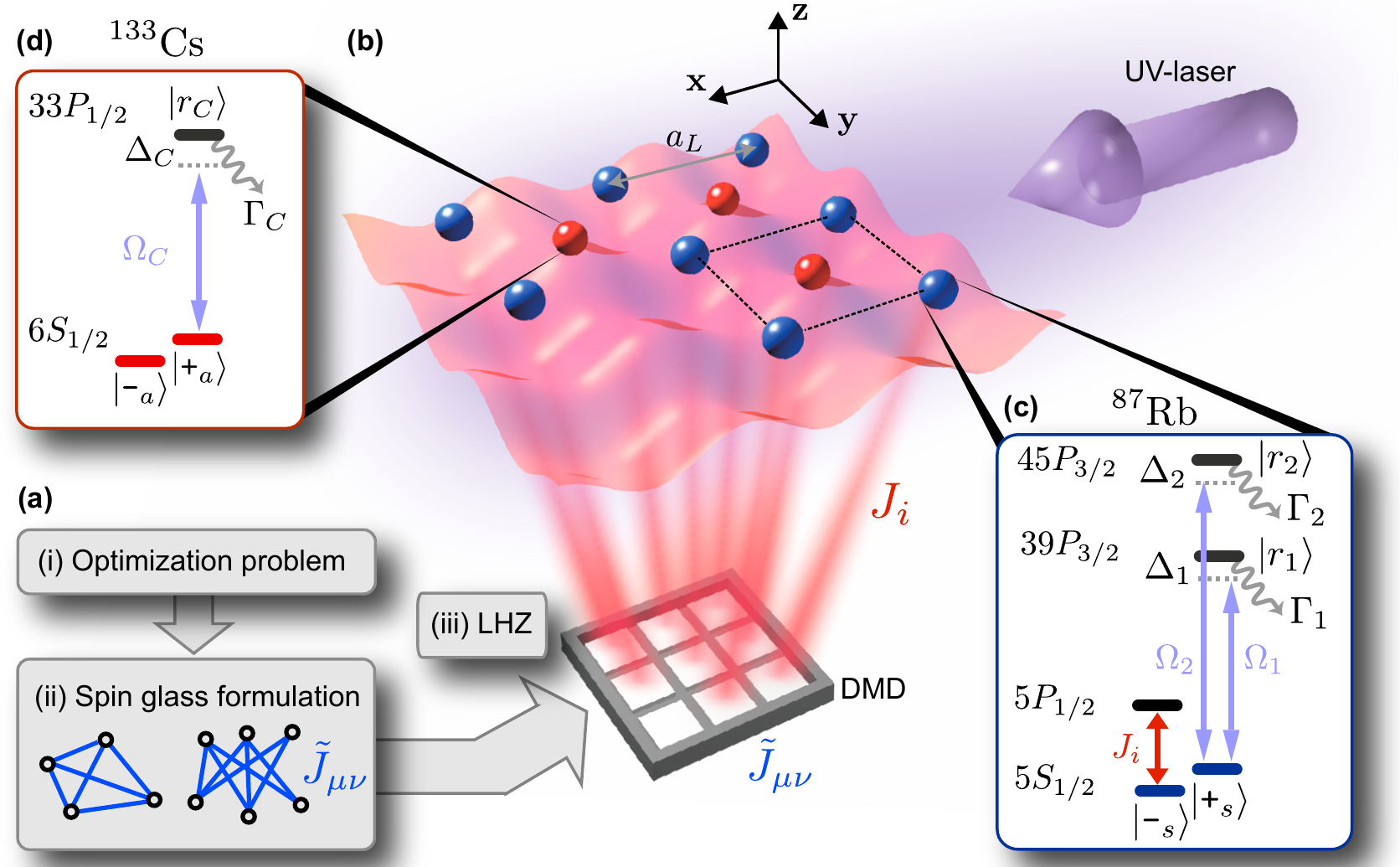}
\caption{\small{(a) The cost function of a general optimization problem in the form of a spin glass with infinite-range interactions $\tilde J_{\mu\nu}$ is encoded in the LHZ architecture in local fields~$J_i$. (b) Rubidium (blue) and Cesium (red) atoms are trapped in a square lattice geometry representing physical and ancilla spins, respectively, where the spin degree of freedom is encoded in two long-lived hyperfine states $|\pl\rangle$ and $|\mi\rangle$. The programmable local fields $J_i$ are induced by AC stark shifts from laser coupling the $\ket{\ds}$ state to low lying $5P$ states using a digital mirror device. The four-body gauge constraints at each plaquette (e.g. the black dotted square) are engineered using off-resonant laser coupling of the $\ket{\us}, \ket{\usa}$ states to Rydberg $P$-states $|r_1\rangle$, $|r_2\rangle$ or $|r_C\rangle$  and require only uniform illumination of the system with UV laser light. }}
\label{fig1}
\end{figure}

Adopting the LHZ architecture~\cite{lechner2015}, the infinite-range spin glass is translated to a lattice spin model, where new {\it physical} spins $\hat \sigma_z^{(i)}$ represent the relative orientation of two {\it logical} spins $ \ls_z^{(\nu)} \ls_z^{(\mu)}$ of Eq.~\eqref{eq:Hlogic}. If two logical spins are aligned in parallel, i.e. $\mid\uparrow\uparrow\rangle$ or $\mid\downarrow\downarrow\rangle$, then the corresponding physical spin is in state $|\pl\rangle$, while if the logical spins are aligned anti-parallel, i.e. $\mid\uparrow\downarrow\rangle$ or $\mid\downarrow\uparrow\rangle$, then the physical spin is in state $|\mi\rangle$. The major advantage of this approach is that the interaction energy of a pair of logical spins can now be implemented with a local field acting on a single physical spin.

A general optimization problem in the LHZ architecture becomes 
\begin{equation}
 \hat H_t = A_t \sum_{i}^K a_i \hat\sigma_x^{(i)} + B_t \sum_{i}^{K}   J_{i}  \hat \sigma_z^{(i)} + C_t  \sum_{\square} \hat H_{\ssq},
\label{eq:Hphys}
\end{equation}  
with new schedule functions $A_t$, $B_t$ and $C_t$ and transverse fields $a_i$. Physical spins are arranged on a square lattice [see blue spheres in Fig.~\ref{fig1}], where the index $i$ labels the entries of the matrix $\tilde J_{\mu\nu}$.
The number of physical spins $K$ equals the number of connections in the original model, which is quadratically larger than in the original problem~\footnote{For example $K=N(N-1)/2$ for all-to-all connected graphs}. This enlarged state space contains states where collections of physical spins encode conflicting relative orientations of the logical spins. These states can be locally identified and energetically penalized by 4-body constraints $H_{\ssq}$ at each plaquette~\scalebox{0.9}{$\square$} of the square lattice, such that at the end of the sweep plaquettes either contain all an even~\cite{lechner2015}, or all an odd~\cite{Rocchetto2016} number of spins in the $\ket{\mi}$ state, thus realising an even or odd parity representation of Eq.~\eqref{eq:Hphys}. This ensures that the final ground state of the LHZ Hamiltonian~\eqref{eq:Hphys} corresponds to the final ground state of the logical Hamiltonian~\eqref{eq:Hlogic}, and thus to the optimal solution of the optimization problem. Importantly, the optimization problem is now encoded in {\it local} fields $\tilde J_{\mu\nu}\rightarrow J_{i}$, corresponding to single particle energy shifts. We show that the above model for a programmable quantum annealer can be emulated in an atomic Rydberg setup, which builds on the remarkable recent advances towards realizing complex spin models with cold atoms in lattices interacting via designed Rydberg-Rydberg interactions~\cite{Schauss2015, Maller2015, Zeiher2016, Jau2016, Labuhn2016}. 

\section{Four-body parity constraints}
The key challenge of implementing $\hat H_{\ssq}$  is resolved with Rydberg atoms by combining (i)~the odd parity representation~\cite{Rocchetto2016}  of  Eq.~\eqref{eq:Hphys} with (ii)~enhanced Rydberg-Rydberg dressing \cite{vanBijnen2015} schemes in a two-species mixture~\cite{Beterov2015, Naegerl}.  In the odd parity representation, the sum of the four spins at each plaquette is either 2 or -2. We introduce a single ancilla qubit $\tau^{\ssq}$ at each plaquette, which can compensate the two associated energies and make odd parity plaquette states degenerate ground states of the constraint Hamiltonian $\hat H_{\ssq}=(\Delta_{\ssq}/4)\hat S_{\ssq}^2$, with stabilizer operators $\hat S_{\ssq}=\sum_{i\in\ssq}\hat \sigma^{(i)}_z+2 \hat \tau_z^{\ssq}$, and energy gap $\Delta_{\ssq}$. This allows to implement the four-body gauge constraints via appropriately designed two-body Ising interactions between physical and ancilla qubits.

\begin{figure}[tb]
\centering
\includegraphics[width= .85 \columnwidth]{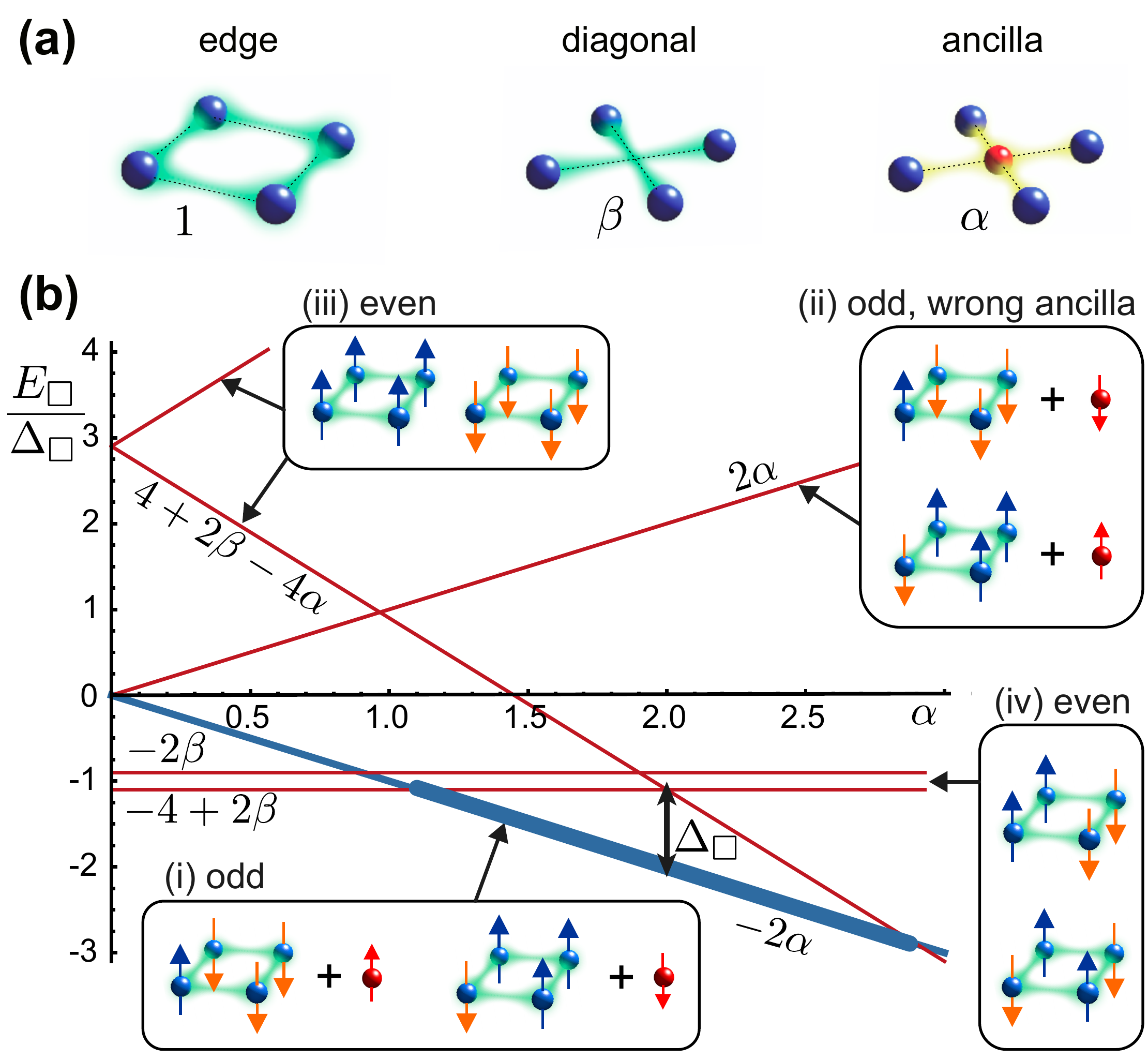}
\caption{\small{(a) Four-body interactions between physical spins  (blue) of the same plaquette are constructed from two-body interactions between physical spins of strength $1$ along the edge of the plaquette (left), interactions of strength $\beta $ along the diagonal (middle) and additional interactions of strength $\alpha$ between an ancilla qubit (red) located at the center of each plaquette and the surrounding physical qubits (right). (b) Eigenenergies $E_{\square}$ of the Hamiltonian of Eq.~\eqref{eq:HCtwobody}, as a function of the physical spin-ancilla interaction strength $\alpha$ for a particular $\beta \lesssim 1$. Odd parity eigenstates with the right (i) or wrong (ii) ancilla orientation have an energy $\pm 2 \alpha$. The maximally polarized states (iii) with all four physical spins up or down have energy $4+2\beta\mp 4\alpha$, while the `spin-ice' states (iv) are independent of the ancilla interaction $\alpha$ and have constant energies $-2\beta$ and $-4+2\beta$. The thick blue line indicates the window of interest where the odd parity states are the ground states of the plaquette Hamiltonian. }}
\label{fig:ancilla}
\end{figure}

Here we consider a more general and robust form of $\hat H_{\ssq}$, consisting of all combinations of two-body  interactions along the edges and diagonals of the plaquette, as well as with the ancilla spin [see Fig.~\ref{fig:ancilla}(a)], of the form
\begin{equation}
\frac{\hat H_{\ssq}}{( \Delta_{\ssq}/2)}=\!\!\!\!\!\!\sum_{i, j \in \mathrm{edges}} \!\!\!\!\!{\hat \sigma}^{(i)}_z {\hat \sigma}^{(j)}_z+\beta\!\!\!\!\!  \sum_{i, j \in \mathrm{diag.}} \!\!\!{\hat \sigma}^{(i)}_z {\hat \sigma}^{(j)}_z + \alpha \hat \tau_z^{\ssq} \sum_{i\in\square} {\hat \sigma}^{(i)}_z,
\label{eq:HCtwobody}
\end{equation}
where $\alpha$ and $\beta$ are relative interaction strengths compared to spin interactions along the plaquette edge.  The energy spectrum $E_{\ssq}$ of a single plaquette Hamiltonian is shown in Fig.~\ref{fig:ancilla}(b), as a function of the parameters $\alpha$ and $\beta$. Importantly, there exists a parameter regime $0 < 2 - \beta < \alpha < 2 + 2\beta$ with $0 < \beta < 1$, where the odd parity states are degenerate and have a lower energy than the even parity states.  Since the precise value of the gap  in Fig.~\ref{fig:ancilla} is not relevant as long as it exceeds all other energy scales, $\hat{H}_{\ssq}$ is quite robust against small variations in interaction strengths, and the parameters $\alpha, \beta$ need not be fine-tuned.

\section{Rydberg implementation}
In the Rydberg quantum annealer illustrated in Fig.~\ref{fig1}, qubits are encoded in two long-lived hyperfine ground states $\ket{\us}$, $\ket{\ds}$  of $^{87}$Rb and  $\ket{\mi_a}$, $\ket{\pl_a}$ of $^{133}$Cs, representing physical and ancilla spins, respectively. These states are trapped in a deep optical or magnetic square lattice with unity filling~\cite{Barredo2016, Endres2016} and frozen motion~\cite{vanBijnen2015,Glaetzle2014,Glaetzle2015,Beterov2015}. In particular, we choose  the $|F=1,m_F=1\rangle$ and $|F=2,m_F=2\rangle$ hyperfine states of the $5^2S_{1/2}$ ground state manifold of  $^{87}$Rb and the $|F=4,m_F=4\rangle$ and $|F=3,m_F=3\rangle$ hyperfine states of the $6^2S_{1/2}$ ground state manifold of $^{133}$Cs. Choosing mixtures of Rb and Cs has the advantage that unwanted cross-talk will be strongly suppressed compared to a single species implementation. The first term of Eq.~\eqref{eq:Hphys} can be realized with a coherent driving field of amplitude $a_i$ coupling the two physical spins, written in a rotating frame. The second term is obtained using single-particle AC-Stark shifts from off-resonant laser coupling of the $\ket{\ds}$ spin state to low-lying $|e\rangle=|5^2P\rangle$ states using a digital micro-mirror device~\cite{Fukuhara2013}.

To implement the two-body interactions of Eq.~\eqref{eq:HCtwobody} we turn to the technique of Rydberg dressing~\cite{Pupillo2010, Henkel2010, Honer2010, Rolston2010}, where off-resonant laser light weakly admixes some Rydberg character into a ground state level.  
Specifically, we propose to couple the $\ket{\us}$ and $\ket{\usa}$ states of Rb and Cs using single photon transitions to Rydberg $P$ states \cite{Zeiher2016, Ott2015, Faoro2015, Jau2016}. 
For large laser detunings the Rydberg dressing acts as a perturbation. Thus the states $\ket{\us}, \ket{\usa}$ predominantly retain their ground state character and remain trapped.

\begin{figure}[tb]
\centering
\includegraphics[width=\columnwidth]{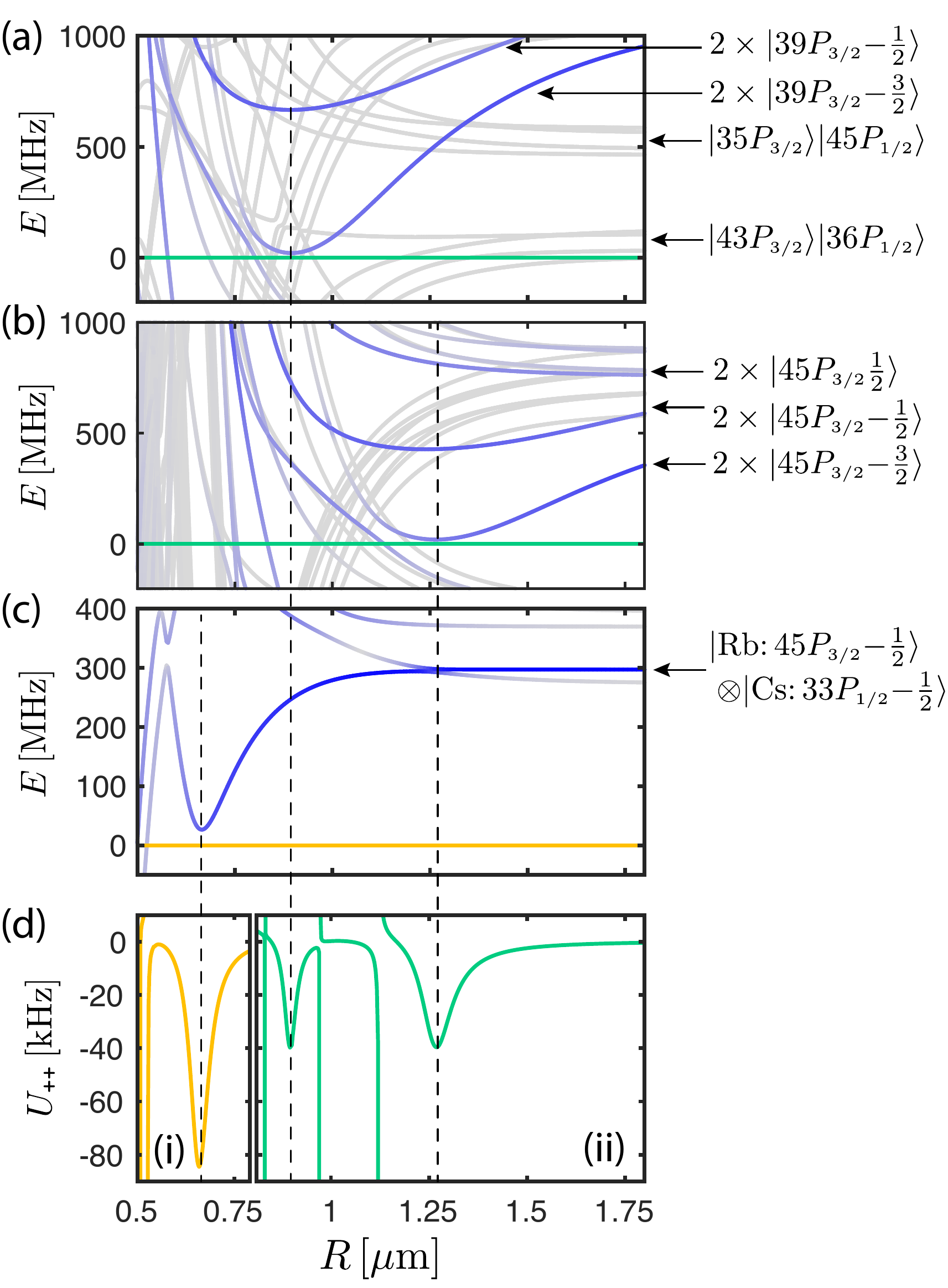}
\caption{\small{Rydberg-Rydberg interaction energies $E$ around the (a) $|39P_{3/2},39P_{3/2}\rangle$ and (b) $|45P_{3/2},45P_{3/2}\rangle$ states of $^{87}$Rb and (c) the mixed $|33P_{1/2},45P_{3/2}\rangle$ Rydberg state of $^{133}$Cs and $^{87}$Rb in a magnetic field of $B_z = 26$~G along the $z$ axis. Due to a close-by F\"orster resonance the pair potentials develops a local minimum. The Rydberg states are chosen such that the position of the minima of these wells are commensurate with the square lattice geometry. The intensity of the blue coloring indicates the overlap with the targeted Rydberg states. Panel~(d) shows the resulting interaction potential $U_{\pl\pl}$ between two Rydberg-dressed (i) Rb-Cs (yellow) and (ii) Rb-Rb (green) ground state atoms. Exciting close to the minima of the interaction potential (vertical dashed lines) yields drastically enhanced peaked-like interactions at $a_L/\sqrt{2}$, $a_L$ and $\sqrt{2}a_L$ thereby realizing the required plaquette interaction of Eq.~\eqref{eq:HCtwobody} for $\alpha=2$ and $\beta=1$. }}
\label{fig:setup}
\end{figure}

Interactions between two spins $i$ and $j$ arise as spatially dependent light shifts $\Uu$ and $\Ua$ of the dressed  pair states $\ket{\us \us}$ and $\ket{\us \usa}$, respectively \footnote{In the perturbative limit, higher order processes involving three or more particles are negligible.}. These pair states are coupled via two photon excitations to doubly excited Rydberg states. Due to dipole-dipole interactions the energies of the doubly excited Rydberg states vary strongly as a function of the relative position $R_{ij}$ of the atoms, where the level shifts exceed typical laser detunings even at micrometer distances. Figure~\ref{fig:setup} shows the Rydberg pair energies of (a)~the $2\times \ket{39 P_{3/2}}$ and (b)~the $2\times\ket{45 P_{3/2}}$ Rydberg states of $^{87}$Rb and (c)~the mixed $|{\rm Cs\!:}33P_{1/2}, {\rm Rb\!:}45P_{3/2}\rangle$ Rydberg state, which show potential wells as a function of the relative distance due to the vicinity of a F\"orster resonance~\cite{vanBijnen2015}. Tuning the (two-photon) detuning of the dressing laser close to a minimum of a potential well results in strongly enhanced ground-state light shifts peaked at the position of the potential minima. The particular Rydberg states are chosen such that the dressed ground state potentials $U_{\pl\pl}^{(s)}$ and $U_{\pl\pl}^{(a)}$ plotted in Fig.~\ref{fig:setup}(d) show peaks at distances $a_L/\sqrt{2}$, $a_L$ and $\sqrt{2}a_L$, commensurate with the square lattice geometry. 

\section{Results}
The final spin-spin interactions between atoms $i$ and $j$ following from the light shifts are
\begin{equation}
\hat H_{\rm i}
=\frac{1}{4}\!\sum_{i,j} \left[U_{\pl\pl}^{(s)}(R_{ij}){\hat \sigma}_z^{(i)}{\hat \sigma}_z^{(j)}\!\!+U_{\pl\pl}^{(a)}(R_{ij}){\hat \sigma}_z^{(i)}{\hat \tau}_z^{(j)} \right],
\label{eq:Hint}
\end{equation}
apart from additional single-particle corrections to the local fields. The height of the two peaks of $U_{\pl\pl}^{(s)}$ at $a_L$ and $\sqrt{2}a_L$ for Rb-Rb  [green line in Fig.~\ref{fig:setup}(d)] and of $U_{\pl\pl}^{(a)}$ at $a_L/\sqrt{2}$ for Rb-Cs  [yellow line in Fig.~\ref{fig:setup}(d)] can be tuned by varying the Rabi frequencies and detunings. In particular, we choose Rabi frequencies \mbox{$\Omega_1 = \Omega_2 = 2\pi \times35\; \mathrm{MHz}$}, \mbox{$\Omega_C = 2\pi \times 13\; \mathrm{MHz}$}, and detunings \mbox{$\Delta_{1} = - 2\pi \times 618\; \mathrm{MHz}$}, \mbox{$\Delta_{2} = - 2\pi \times 373\; \mathrm{MHz}$} and \mbox{$\Delta_C = 2\pi \times 83\; \mathrm{MHz}$} which leads to light-shifts of \mbox{$\Uu(a_L) = \Uu(\sqrt{2}a_L) = - 2 \pi \times 40\; \mathrm{kHz}$} for Rb-Rb and \mbox{$\Ua(a_L/\sqrt{2}) = - 2\pi \times  80 \;\mathrm{kHz}=2\Uu(a_L)$} for Rb-Cs (see supplementary information). 
We note that an external magnetic field and slight vertical offset of the Cs atoms is used to obtain these numbers. 

All interactions are at least two orders of magnitude smaller at lattice distances that do not belong to the plaquette, thus minimizing unwanted cross-talk. This allows to restrict the sum in Eq.~\eqref{eq:Hint} to atoms belonging to the same plaquette, thus realizing $\hat H_{\ssq}$ of Eq.~\eqref{eq:HCtwobody} for the optimal parameters $\alpha=2$ and $\beta=1$. For the above system parameters, we obtain a final energy gap $\Delta_{\ssq}= - 2 \pi \times 20\;\mathrm{kHz}$~\footnote{Note that the energy gap is negative, which can be easily accounted for by a sign change of all local fields and making the annealer adiabatically follow the maximum energy state, instead of the minimum.}. Due to the finite lifetime of the Rydberg states, the dressed ground state interactions come at a cost of an effective decoherence rate $1/\tau_0$ for each qubit. However, since there is only a small Rydberg component admixed, the effective decay rate is also only a correspondingly small fraction of the bare Rydberg decay rate. Ultimately, the figure of merit for fully coherent operation of the quantum annealer is the ratio of the attained interaction strength versus the effective decay rate. In the enhanced dressing scheme this ratio becomes particularly favorable and is of the order of $|\Delta_{\ssq}|\tau_{0} \approx10^3$ for the system parameters above (see supplementary information).

\begin{figure}[tb]
\centering
\includegraphics[width=8cm]{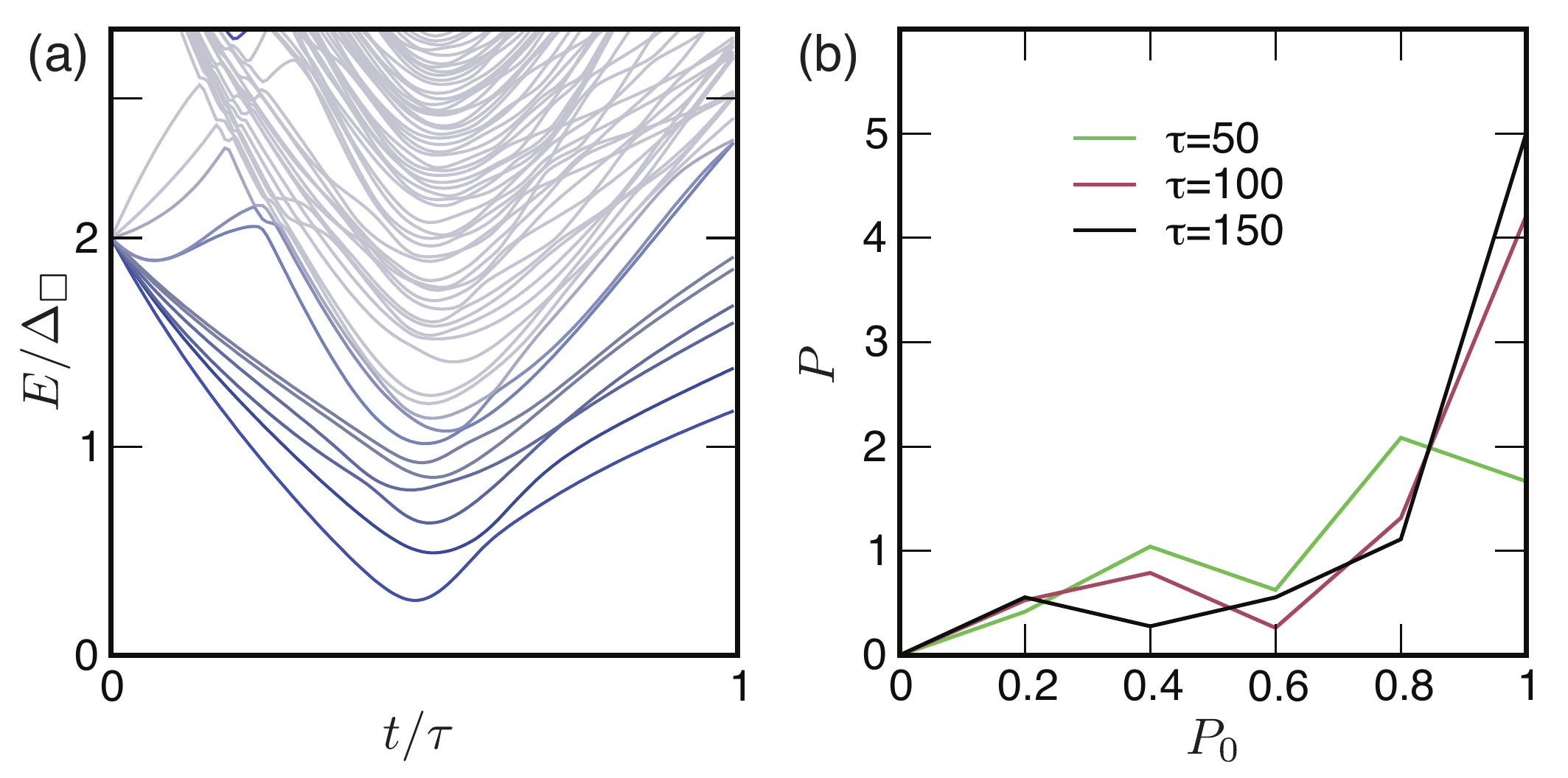}
\caption{\small{(a) Illustration of the time-dependent spectrum for the minimal instance shown in Fig. \ref{fig1}. (b) Histogram of the success probability, i.e. the probability $P_0$ to populate the ground state at final time $\tau$ for different sweep times. }}
\label{fig:fig4}
\end{figure}

Using the above potentials we demonstrate numerically the feasibility of the Rydberg annealer for the minimal instance (see Fig.~\ref{fig1}) with $8$ qubits and $3$ ancillas. Fig.~\ref{fig:fig4} depicts the time dependent spectrum in reduced units. The time-dependent spectrum for instance of Hamiltonian Eq.~(\ref{eq:Hphys}) for random $| J_i/\Delta_{\ssq}| < 1$. The sweep functions $A_t$, $B_t$ and $C_t$ are simple linear functions. Note, that the efficiency can be considerably increased by adopting non-linear sweep functions. In Fig.~\ref{fig:fig4} all energies are given relative to the ground state energy. The pronounced minimal gap is an order of magnitude smaller than the gap in the final state. Figure~\ref{fig:fig4}(b) shows the histogram of the success probability $P_0=|\bra{\psi(\tau)} \psi_{\rm gs} \rangle|^2$, defined as the overlap of the final state $\psi$ with the ground state $\psi_{\rm gs}$, averaged over $N_r=40$ random instances for different sweep times far below the decoherence times $\tau < \tau_{\textrm{0}}/K$. For the fastest switching time $|\Delta_{\ssq}|\tau = 50$ the average success probability is $75\%$ and approaches unity for slower sweeps. 

\section{Conclusions} 
The proposed implementation of a quantum annealer with ultracold Rydberg atoms in optical lattices provides a new platform for adiabatic quantum computing, featuring a highly controllable environment to explore the many-body adiabatic passage, the role of entanglement and effects of decoherence during the annealing sweep. The large lifetimes of Rydberg dressed atoms enable coherent quantum annealing as an alternative to the current paradigm of quantum enhanced thermal annealing~\cite{amin2015}. We anticipate that due to the coherent evolution the number of spins in future experiments can readily be extended well beyond the minimal example presented here, by using shorter annealing cycles with many repetitions~\cite{Knysh2016}, or by employing counter-diabatic driving schemes that  could greatly increase the attained fidelities~\cite{Sels2016}. Finally, our proposal allows to realize atomic quantum simulators of arbitrary infinite-range Ising spin glass models (see e.g. Refs. in \cite{KATZGRABERSG}), and the combination of multi-color Rydberg-dressed interactions with two-species mixtures has applications in realizing $\mathbb{Z}_2$ lattice gauge theories beyond the present example~\cite{Kitaev2003}.

We acknowledge discussions with C.~Gross, H.~C.~N\"agerl, M.~Saffman and J.~Zeiher. This work is supported by the Austrian Science Fund SFB FoQuS (FWF Project No. F4016-N23), the European Research Council (ERC) Synergy Grant UQUAM and the EU H2020 FET Proactive project RySQ. WL acknowledges funding from the Hauser-Raspe Foundation. 

\begin{appendix}
\widetext
\section{Numerical example using LHZ}\label{sec:Num}
A general quantum annealing problem with an infinite-range spin glass Hamiltonian consisting of $N$ logical spins $\tilde\sigma$ with $K$ connections has the form
\begin{equation}
\tilde H_t^{(\rm log)}= \tilde A_t \sum_{\nu=1}^N \tilde a_\nu \tilde\sigma_x^{(\nu)} + \tilde B_t \sum_{\nu<\mu}^{N}\tilde J_{\mu\nu}  \tilde\sigma_z^{(\nu)} \tilde\sigma_z^{(\mu)}
\label{eq:Hlogic}
\end{equation}
with scheduling functions $\tilde A_t$ and $\tilde B_t$, local transverse fields $\tilde a_\nu$ and programmable infinite-range interactions $\tilde J_{\mu\nu}$. 

\begin{figure}[tb]
\centering
\includegraphics[width= .3 \textwidth]{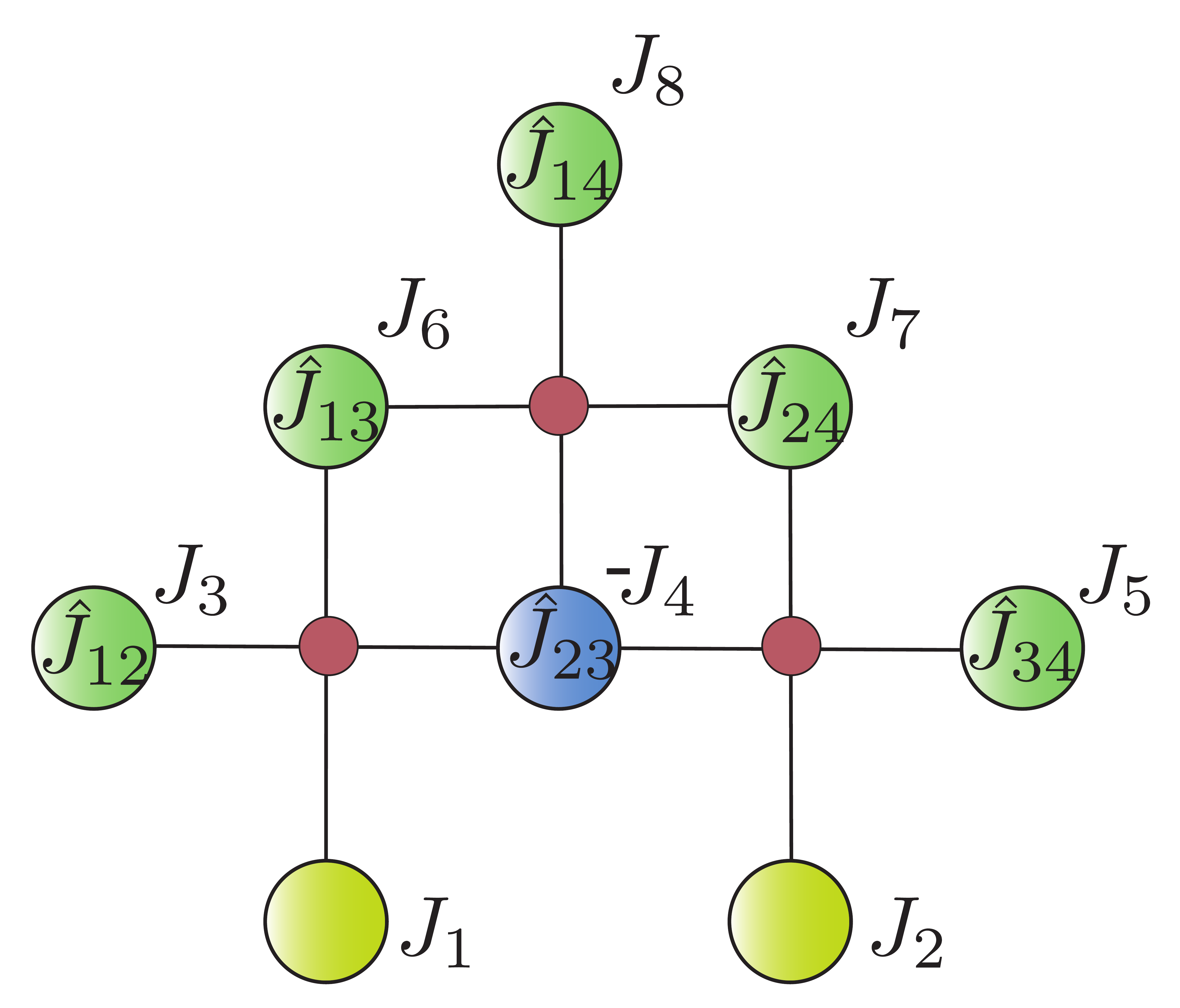}
\caption{The minimal instance shown in Fig.~1 in the main text corresponds to a fully connected spin glass with $N=4$ spins. In the LHZ architecture, the entries of the interaction matrix $\tilde J_{\mu\nu}$ in Eq.~\eqref{eq:Hlogic} translate to local fields ${J}_i$. Here, the index $i$ label local fields ordered from bottom to top. The bottom row of local fields are fixed to large positive numbers $J_{1}=J_{2} = 5|\Delta_{\square}|$.  The minimal instance discussed here consists of $N=8$ spins with $3$ ancilla spins accounting for the three 4-body plaquette constraints (red dots). }
\label{SIfig2}
\end{figure}

Using the LHZ architecture it can be mapped  on a spin model
\begin{equation}
\hat H_t ^{(\rm LHZ)}= A_t \sum_{i}^K a_i\hat \sigma_x^{(i)} + B_t \sum_{i}^{K}  {J}_{i}' \hat \sigma_z^{(i)} + C_t \sum_{\square}\Delta_{\ssq}\prod_{i\in\square}\hat \sigma_z^{(i)},
\label{eq:Hphys}
\end{equation}  
with $K$ physical spins $\hat \sigma$, arranged on a square lattice (green circles in Fig.~\ref{SIfig2}), and problem independent 4-body interactions between spins belonging to the same plaquette $\square$ of the square lattice (red dots in Fig.~\ref{SIfig2}). Here, $a_i$ are transverse local fields and $\Delta_{\ssq}$ is the four-body interaction strength (which, for simplicity, we assume to be equal for all spins and plaquettes, respectively), and $C_t$ is the  scheduling function of the constraints. In the LHZ architecture the programmable interaction matrix $\tilde J_{\mu\nu}$ is translated to programmable single-particle energy shifts ${J}'_{i}$ which correspond to the entries of the matrix $\tilde J_{\mu\nu}$. 

In an odd parity representation the 4-body interactions are resolved by introducing an ancilla qubit $\hat \tau^{\ssq}$ in the middle of each plaquette with fine-tuned 2-body interactions of the form
\begin{equation}
\hat H_t ^{(\rm odd)}= A_t\left( \sum_{i}^K a_i  \hat \sigma_x^{(i)}+\sum_{\square} a_{\ssq} \hat \tau_x^{\ssq}\right) + B_t \sum_{i}^{K}  {J}_{i} \hat \sigma_z^{(i)} + C_t \sum_{\square}\Delta_{\ssq}\left(\sum_{i\in\square}\hat \sigma_z^{(i)}+2\hat \tau_z^{\ssq}\right)^2,
\label{eq:Hodd}
\end{equation} 
where the local fields $J_i$ are the entries of the matrix $(-1)^{\mu(\nu-\mu)}\tilde J_{\mu\nu}$, and $a_{\ssq}$ is a transverse local field driving the ancilla spins. 

In the following we illustrate the annealing sweep and the time-dependent spectrum of Eq.~\eqref{eq:Hodd} for the minimal instance of 8 logical qubits (Rubidium atoms) and 3 ancilla qubits (Cesium atoms) which makes a total of 11 qubits arranged on three plaquettes illustrated in Fig.~1 of the main text and Fig.~\ref{SIfig2} of the supplemental material. This setup corresponds to 4 all-to-all connected logical qubits in Eq.~\eqref{eq:Hlogic}.

The order of the indices of the local fields is from bottom to top (e.g. $\tilde J_{12} \rightarrow {J}_3$). In the minimal instance depicted in Fig. \ref{SIfig2} these are the three plaquettes formed by physical qubits $(1,3,4,6)$, $(2,4,5,7)$ and $(6,4,7,8)$. In the odd parity scheme, the phase factor $(-1)^{\mu(\nu-\mu)}$ will flip the sign of the fourth local field, i.e.~$J'_4=-J_4$, such that the parity of all plaquettes is odd. 

We demonstrate the feasibility of quantum annealing for this small instance numerically calculating the success probability $P_0$ for constant $\Delta_{\ssq}$ and $a_i$. This is the overlap of the wave function at final time $T$ with the ground state wavefunction $\psi_{\rm gs}$
\begin{equation}
P_0=|\bra{\psi(T)} \psi_{\rm gs} \rangle|^2.
\end{equation}
The statistics of $P_0$ is obtained by solving the time-dependent Schr\"odinger equation with Hamiltonian~\eqref{eq:Hodd}, using the Rydberg interaction potentials given in the main text and in Sec.~\ref{sec:Ryd} of the supplemental information, with exact diagonalization and explicitly including the dynamics of the ancilla qubits.  Statistics is taken from $N_i=40$ random instances of an interaction matrix with $J_{i}/|\Delta_{\ssq}| \in [{-0.5,0.5}]$. The schedule functions $A_t = T-t$ and $B_t=C_t=t$ both interpolate linearly from 0 to $T$. Note, that this linear ramp is a pessimistic toy model and more sophisticated choices of the schedule function could considerably increase the success probability. The bottom two qubits $i=1$ and $i=2$ fixing the gauge constraints in the parity architecture are explicitly part of the dynamics and subject to strong local field $J_1=J_2=5|\Delta_{\ssq}|$. The total time is $|\Delta_{\ssq}|T = 50,100,150$, which is significantly shorter than the expected single particle coherence time of the Rydberg states of Rubidium and Cesium which is $|\Delta_{\ssq}|\tau_{0} \approx 10^3$. The results shown in Fig.~3 in the main text show that for this system size the overlap with the ground state wave function is between $70\%$ and $90\%$ for the given parameters.

\section{Rydberg dressed interactions}\label{sec:Ryd}

\subsection{Rydberg-Rydberg potentials}
\label{sec:RydRbCs}
In this section, we outline the calculation of the Rydberg-Rydberg interaction potentials that underly the effective interactions described in the main text. Throughout the discussion we set $\hbar = 1$ for notational clarity.

The Rydberg states $\ket{r} = \ket{n(\ell s)j m_j}$ of the valence electron with spin $s=1/2$ of alkali atoms, such as rubidium or cesium, are well described by their principal quantum number $n$, orbital angular momentum $\ell$, and total angular momentum $j$ with its projection $m_j$ along the quantization axis. 
The states $\ket{r}$ are eigenstates of a single particle Hamiltonian 
\begin{equation}
\hat{H}_A=\sum_r \left(E_r^{(0)}+\Delta E_{m_j}\right)|r\rangle\langle r|.
\end{equation}
The bare atomic energies $E_r^{(0)}=-E_{\rm Ryd}/(n-\delta_{\ell,j})^{2}$ are different for rubidium and cesium atoms, and are determined by their respective quantum defects $\delta_{\ell,j}$, see e.g. Refs. \cite{Lorenzen1984,Li2003} and \cite{Weber1987}, respectively. We include level shifts $\Delta E_{m_j}=\mu_B g_j B_z m_j $ due to a magnetic field $\mathbf{B}=B_z\mathbf{\hat z}$ which lifts the Zeeman degeneracy and sets the quantization axis along the $z$-axis. Here, $\mu_B = 1.4\;h$~MHz/G is the Bohr magneton and $g_j$ is the Lande factor for the Rydberg level.

The Hamiltonian describing the internal-state dynamics of two particles separated by some distance $\Rvec=(R,\vartheta,\varphi)$, to good approximation, given by
\begin{equation}
\hat{H} = \hat{H}_A^{(1)} \otimes \hat{I} + \hat{I} \otimes \hat{H}_A^{(2)} +\hat{V}_{dd}(\Rvec),
\label{eq:Hpair}
\end{equation}
with the first term of each operator product acting on the first particle, and the second term acting on the second. The single particle atomic Hamiltonians $\hat{H}_A^{(1)}$ and  $\hat{H}_A^{(2)}$ can describe rubidium or cesium atoms.
The third term corresponds to the interactions coupling the internal states of the two particles. For the typical distances considered in this work these interaction couplings are predominantly of dipole-dipole form
\begin{equation}
\begin{split}
\hat{V}_{dd}(\Rvec) &= \frac{{\bf d}_1\cdot {\bf d}_2}{R^3} - \frac{3({\bf d}_1\cdot {\bf R})({\bf d}_2\cdot {\bf R})}{R^5}
=-{ \sqrt{\frac{24 \pi}{5}} } \frac{1}{R^{3}} \sum_{\mu,\nu}C_{\mu,\nu;\mu+\nu}^{1,1;2}
 Y_{2}^{\mu+\nu}(\vartheta,\varphi)^* d^{(1)}_{\mu}d^{(2)}_{\nu}.
 \end{split}
\end{equation}
Here, $\hat{\bf d}_{1}$ and $\hat{\bf d}_{2}$ are the dipole transition operators for atom $1$ and atom $2$, respectively, with spherical components $d_{\mu}^{(1)}$ and $d_{\nu}^{(2)}$ and $(R,\vartheta,\varphi)$ are the spherical components of the relative vector. With $C$ and $Y$ we denote Clebsch-Gordan coefficients and spherical harmonics, respectively.

We proceed by selecting a large basis set of pair states
$|rr'\rangle = |n (\ell s) j m_j\rangle \otimes |n' (\ell's') j' m_j'\rangle$, which are product states of eigenstates of the single particle Hamiltonians $\hat{H}_A^{(1)}$ and $\hat{H}_A^{(2)}$, with energies $E_{rr^\prime}$ given by the sum of the corresponding single-atom energies. In this basis, the pair Hamiltonian of Eq.~\eqref{eq:Hpair} turns into a large but sparse matrix, 
with off-diagonal matrix elements
\begin{equation}
\begin{split}
\langle r_i,r_j|&V_{\rm dd}(\mathbf{R})|r_k,r_l\rangle=\mathcal{R}_{i,k}\mathcal{R}_{j,4}\\
\times&(-)^{s-m_i}{[\ell_i][j_i][\ell_k][j_k]}\ws{\ell_i,\ell_k, 1,j_k,j_i,s}\wt{\ell_k,1,\ell_i,0,0,0}\\
\times&(-)^{s-m_j}{[\ell_j][j_j][\ell_l][j_l]}\ws{\ell_j,\ell_l, 1,j_l,j_j,s}\wt{\ell_l,1,\ell_j,0,0,0}\\
\times&\left[-\sqrt{\frac{24 \pi}{5}} \sum_{\mu,\nu}C_{\mu,\nu;\mu+\nu}^{1,1;2} \wt{j_k,1,j_i,m_k,\mu,-m_i}\wt{j_l,1,j_j,m_l,\nu,-m_j}
 Y_{2}^{\mu+\nu}(\vartheta,\varphi)^*\right],
\end{split}
\end{equation}
with $\mathcal{R}_{i,k}=\langle r_i||r||r_k\rangle$  the radial integral and the abbreviation $[x]=\sqrt{2x+1}$. The matrix is block diagonal, coupling only $\ell$ to $\ell\pm 1$ states and for $\vartheta=\pi/2$ the total magnetic quantum number $\Delta M=m_1+m_2$ can change by 0 or $\pm 2$.

We diagonalize $\hat H$ numerically for a range of distances $R$ for fixed $\vartheta=\pi/2$. The basis set is chosen sufficiently large, containing $\sim 10^4$ states, to ensure convergence of eigenstates and eigenvalues down to distances of $R\sim 0.5 \mu \mathrm{m}$. The diagonalization procedure yields distance dependent molecular eigenenergies $E_\mu({ R})$ depending only on the radial distance $R$, which are the interaction potentials plotted in Fig. 3(a)-(c) in the main text. Simultaneously the corresponding molecular eigenstates $\ket{\mu({\bf R})}$ are computed,
\begin{equation}\label{EqMolecularEigenstates}
\ket{\mu({\bf R})} = \sum_{rr'} c^{(\mu)}_{rr'}({\bf R}) \ket{rr'},
\end{equation}
which are superpositions of pair product states $\ket{rr'}$ with coefficients $c^{(\mu)}_{rr'}({\bf R})$ depending on $(R,\vartheta,\varphi)$. The coloring of the curves in Fig. 3(a)-(c) of the main text is indicative of the overlap with the laser targeted Rydberg state, $|\langle r_{\laserlabel1}r_{\laserlabel2} | \mu(\Rvec) \rangle|^2 = |c^{(\mu)}_{r\laserlabel1, r\laserlabel2}(\Rvec)|^2$, with $\laserlabel1, \laserlabel2 = $\ \{`1',\  `2',\  `C'\} defining the particular Rydberg states [see e.g. Fig.~1 in the main text].

\subsection{Rydberg dressing potentials}
\label{sec:RydDress}

Having obtained the molecular eigenstates $\ket{\mu({\bf R})}$ and their energies $E_\mu(R)$, we can now proceed to calculate the light shifts of the ground state levels resulting from laser coupling to the excited state manifold.

The laser couplings are characterized by a Rabi frequency, $\Omega_\laserlabel$, and detuning from a targeted Rydberg level, $\Delta_\laserlabel$. The subscript $\laserlabel = $\ \{`1',\  `2',\  `C'\} indexes the three distinct laser couplings discussed in the main text: 
\begin{itemize}
\item $\laserlabel = $\ \lq1\rq \ pertains to the laser coupling the $\ket{\us} = \ket{F = 2, m_F = -2}$ hyperfine ground state of Rb to the $\ket{r_1} = \ket{39P_{3/2}, m_J = -1/2}$ Rydberg state,
\item $\laserlabel = $\ \lq2\rq \ indexes the laser coupling of the $\ket{\us}$ state to the $\ket{r_2} = \ket{45P_{3/2}, m_J = -1/2}$ Rydberg state, 
\item $\laserlabel =  $\ \lq C\rq \ refers to the laser coupling of the $\ket{\usa} = \ket{F = 4, m_F = -4}$ hyperfine ground state of an ancilla Cs atom to the $\ket{r_C} = \ket{33P_{1/2}, m_J = -1/2}$ Rydberg state.
\end{itemize}
All lasers propagate in the $xy$-plane, with linear polarization along the $z$-axis coinciding with the quantization axis and magnetic field direction. This geometry is chosen such that the total system and resulting interaction potentials are rotationally symmetric along the $z$-axis, and in particular the energies of the plaquette configurations are invariant under rotation and mirroring operations.

In the Rydberg dressing limit, the laser coupling is far off-resonant with $\Omega_\laserlabel \ll |\Delta_\laserlabel |$, such that the effect of the laser coupling is perturbative with an associated small parameter $\epsilon = \Omega_\laserlabel / |\Delta_\laserlabel|$. In the following we only consider the laser-coupled ground states $\ket{\pl_s}, \ket{\usa}$, as the uncoupled ground states $\ket{\mi_s}, \ket{\dsa}$ play no role in the interaction calculation. To describe the state of two laser coupled ground state particles, we work in a basis consisting of  pair states $\ket{g_1g_2}$, where both particles are in one of the two ground state $|\pl_s\rangle$ or $|\pl_a\rangle$, i.e. $g_1, g_2 = \pl_s, \usa$. This basis is extended with pair states $\ket{g_1 r_{\laserlabel2}}, \ket{r_{\laserlabel1} g_2}$ where one of the two particles is in the ground state while the other is excited to the laser-targeted Rydberg state, with $\laserlabel1, \laserlabel2 = $\ \{`1',\  `2',\  `C'\}. Due to appropriately chosen laser frequency and polarization we only couple to these targeted Rydberg states. Finally, the basis also contains the molecular states $\ket{\mu(\Rvec)}$ we obtained numerically in the previous section.

The ground pair state $\ket{g_1g_2}$ is defined to have an energy $0$.
The atomic Rydberg states of particle $1$ are defined in a rotating frame corresponding to the laser frequency $\omega_{\laserlabel1}$, such that the near-resonant, laser-targeted Rydberg state $\ket{r_{\laserlabel1}}$ has an energy $-\Delta_{\laserlabel1}$. Similarly, the Rydberg states of particle $2$ have their energy defined relative to the energy $-\Delta_{\laserlabel2}$ of the $\ket{r_{\laserlabel2}}$ state in the rotating frame of the laser transition $\laserlabel2$. The molecular Rydberg states $\ket{\mu(\Rvec)}$ therefore have an energy 
\begin{equation}
\delta^{(\mu)}(R) = E_\mu({ R}) - E_{r_{\laserlabel1}r_{\laserlabel2}}-\Delta_{\laserlabel1} - \Delta_{\laserlabel2}.
\end{equation}
Expressed in the basis described above, the two-particle Hamiltonian thus becomes
\begin{equation}
\hat{H} = -\Delta_{\laserlabel1} \ket{r_{\laserlabel1}g_2}\bra{r_{\laserlabel1}g_2} - \Delta_{\laserlabel2}\ket{g_1r_{\laserlabel2}}\bra{g_1r_{\laserlabel2}} + \sum_\mu \delta^{(\mu)}(R) \ket{\mu(\Rvec)} \bra{\mu(\Rvec)} + \hat{H}_L({\bf R}),\label{Htwop}
\end{equation}
where the operator $\hat{H}_L$ is the laser coupling
\begin{multline}
\hat{H}_L({\bf R}) = \frac{\Omega_{\laserlabel1}}{2} \ket{r_{\laserlabel1}g_2}\bra{g_1g_2} + \frac{\Omega_{\laserlabel2}}{2} \ket{g_1r_{\laserlabel2}}\bra{g_1g_2} + \mathrm{h.c.} \nonumber\\ 
+\sum_\mu \left[ \frac{\Omega_{\laserlabel1}^{(\mu)}({\bf R})}{2}\ket{\mu(\Rvec)}\bra{g_1 r_{\laserlabel2}} + \frac{\Omega_{\laserlabel2}^{(\mu)}({\bf R})}{2}  \ket{\mu(\Rvec)}\bra{r_{\laserlabel1}g_2}\right] + \mathrm{h.c.},
\end{multline}
where the first line contains terms coupling the pair ground state $\ket{g_1 g_2}$ to the single excited pair states, and the second line couples the singly excited states to the molecular states. The effective coupling strength to the molecular states,
\begin{equation}
\Omega^{(\mu)}_{\laserlabel} ({\bf R})= \Omega_{\laserlabel} c^{(\mu)}_{r\laserlabel1, r\laserlabel2}(\Rvec),
\end{equation}
with $\laserlabel = \laserlabel1, \laserlabel2$, is reduced with a factor $c^{(\mu)}_{r\laserlabel1, r\laserlabel2} \leq 1$, and is additionally dependent on the distance.

\begin{figure*}[tb]
\centering
\includegraphics[width= .3 \textwidth]{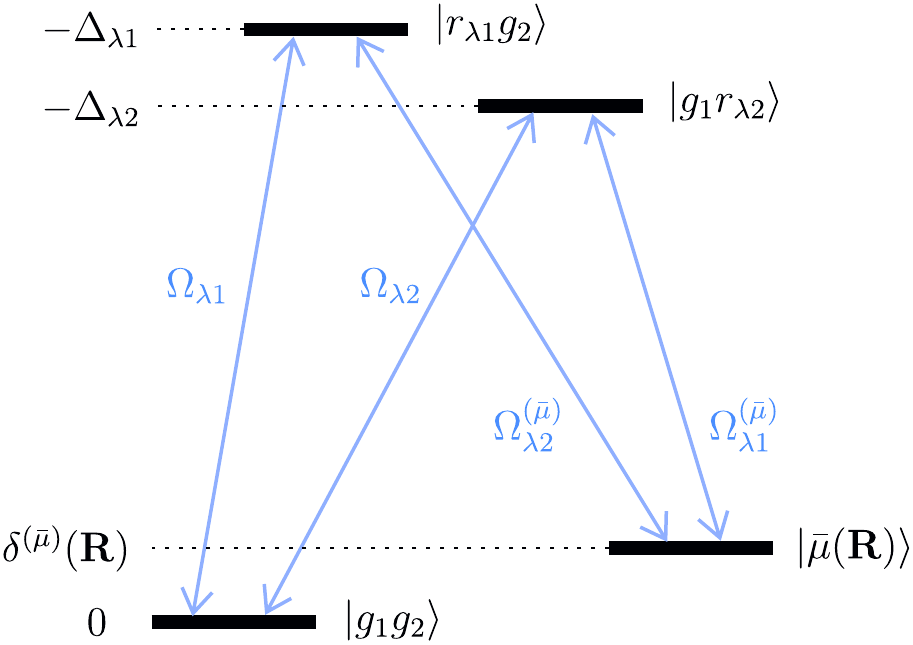}
\caption{\small{Energy levels involved to obtain the dressed ground state potentials. The double ground state $|g_1,g_2\rangle$ is laser-coupled to the $|r_{\lambda 1},g_2\rangle$ or $|g_1,r_{\lambda 2}\rangle$ single-excited Rydberg state with Rabi frequencies $\Omega_{\lambda 1}$ and $\Omega_{\lambda 2}$ and energies $\Delta_{\lambda 1}$ and $\Delta_{\lambda 2}$, respectively. These state can be further excited to doubly-excited Rydberg states $|\bar \mu({\bf R})$ with position dependent Rabi frequencies $\Omega_{\lambda 1}^{(\bar \mu)}$ and $\Omega_{\lambda 2}^{(\bar \mu)}$. Close to a molecular potential well the energy $\delta^{(\bar \mu)}(R)$ can be much smaller than the single particle detunings $\Delta_\lambda$ resulting in an enhanced light shift of the dressed $\ket{g_1, g_2}$ state.
}}
\label{SIfig1}
\end{figure*}

The light shifts of the pair ground state presented in the main text, Fig. 3(d), are calculated by numerically diagonalizing Hamiltonian (\ref{Htwop}). Here, however,  it is instructive to analyze the light shifts perturbatively. In particular, we consider the case where there is one dominant molecular state, denoted  $\mu = \bar\mu$, with significant effective laser coupling strength and lying close to $0$ in energy. This is precisely the situation described in the main text in the vicinity of the point of closest approach of the molecular wells to the pair ground state, as depicted in Fig 3(a)-(c) in the main text. Figure \ref{SIfig1} illustrates the relevant pair states, their energies and the couplings between them.

To second order in the laser coupling, the light shift of the pair ground state is constant and independent of distance, 
\begin{equation}
E^{(2)} = \frac{1}{2}\Omega_{\laserlabel1} \epsilon_{\laserlabel1} + \frac{1}{2}\Omega_{\laserlabel2} \epsilon_{\laserlabel2},
\end{equation}
where we have defined $\epsilon_\laserlabel = \Omega_\laserlabel / 2 \Delta_\laserlabel$.
Thus far, the laser light shift is merely a single particle effect.
Interactions enter in fourth order perturbation theory, when we consider processes involving couplings to the molecular state $\ket{\bar\mu(\Rvec)}$. The resulting contribution to the light shift is (ignoring terms contributing to the single particle light shift)
\begin{equation}
E^{(4)}_{\rm int} = \epsilon_{\bar\mu}^2 \delta^{(\bar\mu)}(R), \label{EqVintEff}
\end{equation}
where we have defined
\begin{equation}
\epsilon_{\bar\mu} = \frac{\Omega_{\laserlabel1}\Omega^{(\bar\mu)}_{\laserlabel2}}{4 \delta^{(\bar\mu)}(R)}\left(\frac{1}{\Delta_{\laserlabel1}} + \frac{1}{\Delta_{\laserlabel2}} \right) = \frac{\Omega_{\laserlabel2}\Omega^{(\bar\mu)}_{\laserlabel1}}{4 \delta^{(\bar\mu)}(R)}\left(\frac{1}{\Delta_{\laserlabel1}} + \frac{1}{\Delta_{\laserlabel2}} \right).
\end{equation}
Clearly, choosing the laser detunings such that $\delta^{(\bar\mu)}(R)$ becomes small, boosts the interaction strength.
The above perturbative expression is valid as long as $\delta^{(\bar\mu)}(R) \gg 
\epsilon_\laserlabel \Omega_{\laserlabel}^{(\bar\mu)}$.
A particular situation where the perturbative treatment breaks down occurs when a molecular state crosses the zero  energy level. At such a point, pairs of Rydberg atoms are resonantly excited by the laser, instead of the intended weak admixture. The system parameters in the main text are chosen such that this situation is avoided, by ensuring that no significant resonances occur at lattice distances. We included checking the 'cross' potentials, i.e. the molecular potentials for the case $\laserlabel1 = $`1', and $\laserlabel2 = $`2'. 

A final quantity of interest is the Rydberg state admixing into the atomic ground states, as this determines the dominant decoherence rate in the system.
Again, the state admixing can be calculated in the pair basis and perturbative limit discussed above. The new dressed pair ground state, denoted $\ket{\widetilde{g_1 g_2}}$, becomes
\begin{equation}
\ket{\widetilde{g_1 g_2}} = \ket{g_1 g_2} + \epsilon_{\laserlabel1}\ket{r_{\laserlabel1} g_2} + \epsilon_{\laserlabel2}\ket{g_1r_{\laserlabel2}} - \epsilon_{\bar\mu}\ket{\bar\mu(\Rvec)} + \mathcal{O}\left(\frac{\Omega_{\laserlabel1, \laserlabel2}^2}{\Delta_{\laserlabel1, \laserlabel2}^2}\right),\label{EqStateMixing}
\end{equation}
where we have ignored the normalization, and used $\delta^{(\bar\mu)}(R) \ll \Delta_{\laserlabel1, \laserlabel2}$ to truncate the expansion after the third term. 
Assuming for simplicity a single decoherence rate $\Gamma$ for all Rydberg states, we see that the second and third term in Eq. (\ref{EqStateMixing}) each introduce an effective decoherence rate $\epsilon_\laserlabel^2 \Gamma$ to the dressed pair state, whereas the third term introduces a decoherence rate $2\epsilon_{\bar\mu}^2 \Gamma$, where the factor $2$ in front stems from the fact that the state $\ket{\bar\mu(\Rvec)}$ has two particles in the excited state. The total decoherence per particle is evidently
\begin{equation}
\Gamma_{\rm eff} = \frac{1}{2}(\epsilon_{\laserlabel1}^2 + \epsilon_{\laserlabel2}^2 + 2 \epsilon_{\bar \mu}^2) \Gamma.\label{EqEffGamma}
\end{equation}

The figure of merit for realizing fully coherent operation of the quantum annealer, i.e. the ratio of interaction strength versus decoherence rate (per particle), is now readily computed from the results obtained above. Using Eqs. (\ref{EqVintEff}) and (\ref{EqEffGamma}), we have that
\begin{equation}
\frac{E^{(4)}_{\rm int}}{\Gamma_{\rm eff}} = \frac{2 \epsilon_{\bar\mu}^2 \delta^{(\bar\mu)}(R)}{(\epsilon_{\laserlabel1}^2 + \epsilon_{\laserlabel2}^2 + 2 \epsilon_{\bar \mu}^2) \Gamma} 
.\label{EqFigMerit}
\end{equation}
For the system parameters employed in the main text we thus obtain:
\begin{itemize}
\item $\laserlabel1 = \laserlabel2 = 1$: dressing to $\ket{r_1} = \ket{39P_{3/2}, m_J = -1/2}$, for which the single particle lifetime is $\tau_1 = 1 /\Gamma_1 = 54 \mu \mathrm{s}$ \cite{Beterov2009}, and at the minimum of the selected potential well $\delta^{(\bar\mu)}(R) = 2.5 \mathrm{MHz}$ and $c^{(\bar\mu)} \simeq 0.32$, leading to a final figure of merit $\frac{E^{(4)}_{\rm int}}{\Gamma_{\rm eff}} \simeq 8.0 \times 10^2$,

\item $\laserlabel1 = \laserlabel2 = 2$: dressing to $\ket{r_2} = \ket{45P_{3/2}, m_J = -1/2}$, with single particle lifetime $\tau_2 = 1 / \Gamma _2= 75 \mu \mathrm{s}$ \cite{Beterov2009}, and $\delta^{(\bar\mu)}(R) = 5.5 \mathrm{MHz}$ and $c^{(\bar\mu)} \simeq 0.28$, leading to a final figure of merit $\frac{E^{(4)}_{\rm int}}{\Gamma_{\rm eff}} \simeq 1.9 \times 10^3$,

\item $\laserlabel1 = 2, \laserlabel2 = C$: simultaneous dressing of Rb to $\ket{r_2} = \ket{45P_{3/2}, m_J = -1/2}$ and Cs to $\ket{r_C} = \ket{33P_{1/2}, m_J = -1/2}$, with averaged single particle lifetime $\tau_C = 1 /  \Gamma_C = 50 \mu \mathrm{s}$ \cite{Beterov2009}, and $\delta^{(\bar\mu)}(R) = 10.6 \mathrm{MHz}$ and $c^{(\bar\mu)} \simeq 0.55$, leading to a final figure of merit $\frac{E^{(4)}_{\rm int}}{\Gamma_{\rm eff}} \simeq 2.1 \times 10^3$.

\end{itemize}
A final point of attention concerns a small vertical offset of the Cs atoms in the $z$-direction, which is necessary for getting an exact match of the potential peaks with the lattice geometry with lattice spacing $a_L = 0.89 \mu \mathrm{m}$. The Rb-Cs potential for the chosen parameters has its peak at $0.66 \mu {\mathrm{m}}$, which is slightly larger than the required $a_L / \sqrt{2} = 0.63 \mu \mathrm{m}$. A vertical offset of the Cs atoms of $\approx 200 \mathrm{nm}$ compared to the plane of the Rb atoms would compensate for this difference.

\end{appendix}


%

\end{document}